\documentclass{ws-procs9x6}
\usepackage{epsfig}
\begin{document}

\title{Neutral long-living kaon and muon system of the Belle II detector}
\author{Timofey Uglov}
\address{Institute for Theoretical and Experimental Physics, \\
Bolshaya Cheremushkinskaya 25,
117218, Moscow,
Russia; \\
National Research Nuclear University MEPhI, \\
Kashirskoye shosse 31, Moscow, 115409, Russia\\
E-mail: uglov@itep.ru
}

\begin{abstract}
The Belle detector operated at KEKb B-factory in 1999-2010 was one of
the most remarkable experiments in the field of elementary particle
physics of the last decades. The Belle successor, Belle II
collaboration, is aimed to operate the Belle II detector at SuperKEKb
factory at 40 times higher luminosity. Increased luminosity imposes
new requirements on the detector elements: they have to survive at
higher radiation levels, to operate at higher loads and at higher
backgrounds. The Belle $K_L$ and muon system based on the resistive
plate chambers (RPC) technology worked well during all data taking
period, however at Belle II environments its performance decreases to
negligible level due to increasing load and high neutron
background. To sustain detector operation it will be replaced by the
new system based on the scintillation strips read-out by silicon
photomultipliers.  The latter technology allows not only to reach time
resolution at level of 1~ns but also perform the amplitude
measurements. Nowadays the production of the new EKLM system's
elements are under way. The assembly at KEK is started this fall.
\end{abstract}


\section{Belle and Belle II}
\label{intro}

In the constellation of the brilliant elementary particle experiments
of the last decades, the B-factories, Belle~\cite{belle} and
BaBar~\cite{babar}, were ones of the most remarkable.  Their results
on CP-violation in B-mesons prove the validity of the KM mechanism,
which inventors, M.~Kobayashi and T.~Maskawa, won the Nobel Prize in
physics in 2008. Belle was successful not only in CP-violation
studies. The huge amount of the data collected by the Belle detector
(more than one billion $B\overline{B}$-pairs) allows to make a number
of discoveries in hadron spectroscopy. More than 10 new
quarkonium-like states were discovered including puzzling X(3872)
state and new charged Z-resonances which do not fit
conventional quarkonium model.

The Belle successor, Belle II experiment~\cite{belle2tdr}, is aimed to
collect statistics sample of 50 ab$^{-1}$ by the end of 2022. Major
accelerator upgrade includes changes of the crossing angle (22~mrad to
83~mrad), introduction of the nano-beam scheme, {\em i.e.}  decreasing
of horizontal emittance ($18$~nm to $3.2$~nm for positrons and
$24$~nm to $4.6$~nm for electrons) and beta-functions at IP
($\beta^*_x/\beta^*_y$ from $1200/5.9$~mm to $\approx30/0.3$~mm),
rising beam currents (from $1.64$~A to $3.6$~A for positrons and from
$1.19$~A to $2.6$~A for electrons). The downside of these changes is
significant rise of backgrounds.  Higher luminosity delivered by the
accelerator requires adequate upgrade of the detector components to
sustain effective operation at new environment. New Belle II detector
have 2-layer DEPFET pixel (PXD) and 4-layer DSSD silicon strip (SVD)
detectors for vertexing. Tracking is done with the upgraded drift
chamber (CDC) with increased size and smaller cells. Particle
identification is done with time of propagation counter (TOP)
combining advantages of the excellent time resolution time-of-fight
system and \v{C}erenkov ring imaging BaBar-like DIRC in barrel region
and aerogel RICH (ARICH) detector with focusing radiator in
endcaps. Electromagnetic calorimeter with new electronics for barrel
and pure CsI crystals instead of CsI(Tl) in endcaps is able to operate
at much higher backgrounds. Muon and $K_L$ system (KLM) is replaced
with new one based on scintillator technique in endcaps and in two
innermost barrel layers.

\section{Belle KLM system}
The $K_L$ and muon detector (KLM) of the Belle detector consists of an
alternating sandwich of $4.7$-cm thick iron plates and active detector
elements located outside of the superconducting solenoid. The
octagonal barrel covers the polar angle range from $45^\circ$ to
$125^\circ$, while the endcaps extend this coverage from $20^\circ$ to
$155^\circ$ . There are 14 detector layers and 14 iron plates in each
endcap. Each layer operates independently. The Belle KLM is based on
resistive plate chambers (RPC) technology. High voltage is distributed
over the two glass electrodes with bulk resistivity of
$\sim5\times10^{12}\,\Omega\cdot\mathrm{cm}$ but transparent to the
fast electric pulses. The throughgoing charged particle ionizes gas
along its track and form streamer discharge in the gas-filled gap
between electrodes. The signal induced on the couple of orthogonal
pick-up strip electrodes is large enough to be discriminated without
amplification.

In the Belle environment the performance of the most of the barrel
layers was about $99\%$ decreasing down to $91\%$ for the innermost
one. For the endcap efficiency ranges between $95\%$ and $76\%$. The
lowest numbers are for the outermost layers where the background is
higher and came mostly from the accelerator tunnel.

\section{Belle II KLM system}

Since at high background loads of Belle II experiment efficiency of the
RPCs drops to zero, new Belle II endcap KLM system (EKLM) utilizes
another technique based on scintillator strips read-out by wavelength
shifting (WLS) fibers. The light is registered by novel devices --
silicon photomultipliers (SiPM)\footnote{Different manufacturers use
  different names for devices of such type, {\em e.g.} CPTA, MPPC, MRS
  APD.}.

\begin{figure}[h!]
\begin{center}
\epsfig{file=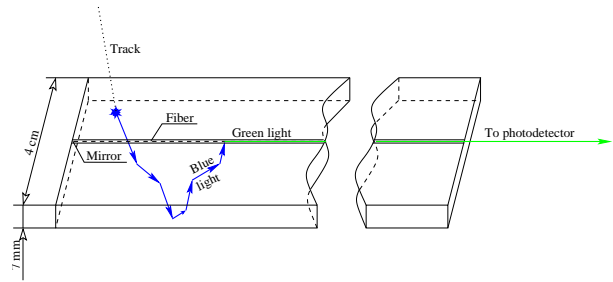,width=0.7\textwidth}
\end{center}
\caption{Scintillator strip with glued WLS fiber.}
\label {strip}
\end{figure}

\begin{figure}[h!]
\begin{center}
\begin{picture}(1684,0)
\put(225,-43){\vector(-2,0){43}}
\put(225,-43){\vector(0,-2){43}}
\put(225,-43){\vector(-1,-1){20}}
\put(225,-43){\vector(-4,1){65}}
\put(225,-43){Preamplifier boards}
\put(255,-105){\vector(-1,-1){45}}
\put(255,-105){\vector(-1,0){47}}
\put(255,-105){\vector(-1,-2){45}}
\put(255,-105){Support grid}
\end{picture}
\epsfig{file=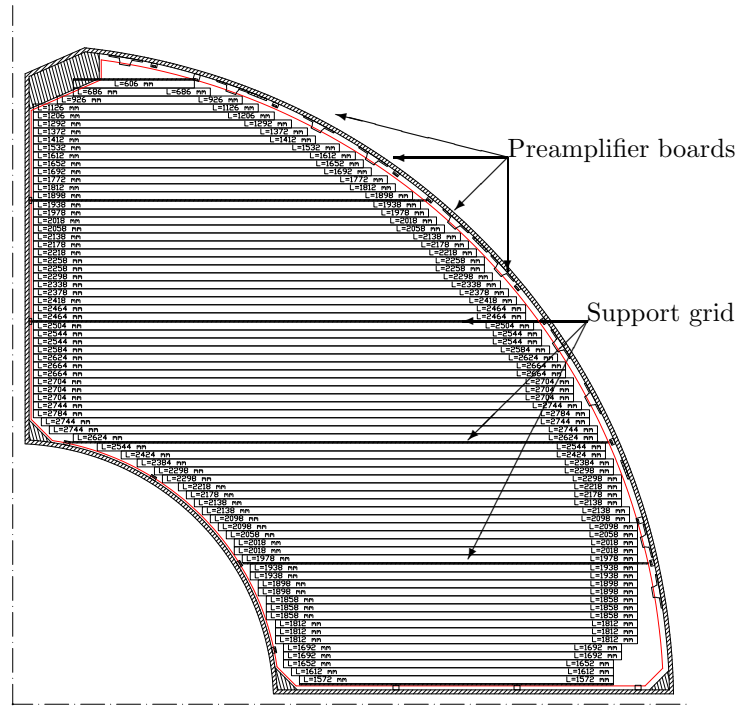,width=0.8\textwidth}
\end{center}
\caption{EKLM sector.}
\label {sector}
\end{figure}

The base element of the new system is scintillator strip of
polystyrene doped with PTP ($1.5\%$) and POPOP ($0.01\%$) produced by
extrusion technique at "Uniplast" enterprise (Vladimir, Russia). It
has a cross-section of $7\times40$~mm$^2$ and length varying from
$606$~mm to $2784$~mm. In total there are 41 different strip lengths.
Diffusive reflective cover of the surface is produced by chemical
etching. A $1.3$ mm wide and $3$ mm deep groove is milled in each
strip to hold the WLS fiber. An optical contact between the
scintillator and Kuraray Y11 WLS fiber put into the groove is provided
by optical gel SL-1 by SUREL (St.-Peterburg, Russia). One WLS fiber
end is connected to a photosensor while the other is mirrored using a
silver-shine paint. The groove is covered by aluminized reflective
tape.  The schematic view of the strip is presented in
Fig.~\ref{strip}. The photosensor is Hamamatsu SiPM
(S10362-13-025CKX). It is pressed against the polished fiber end by
the rubber ring in the connector. SiPMs are connected to the
preamplifiers by twisted-pair cables.  For the sake of efficient
handling strips are glued together into segments of 15 strips.  A
$1.5$~mm plastic plates are glued to the strips at both segment's
sides with double-sided sticky tape.  The whole EKLM system consists
of 112 identical modules. Each module contains 150 strips (10 segments)
in two identical orthogonal planes,  preamplifiers for these
strips and is enclosed into aluminum box. The segments supported by
aluminum I-beams that run along the segment boundaries.  The I-beams
supporting the $x$- and $y$-plane segments are glued into a supporting
grid. The schematic view of the module is shown in Fig.~\ref{sector}.

\begin{figure}[h!]
\begin{center}
\epsfig{file=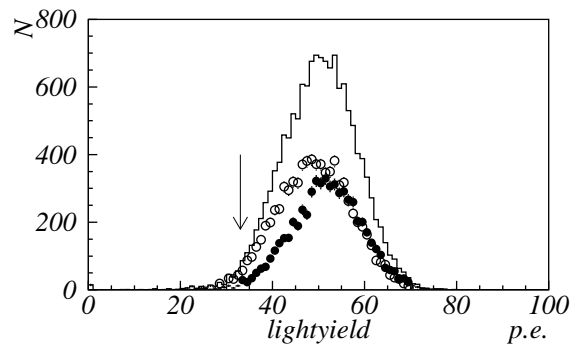,width=0.7\textwidth}
\end{center}
\caption{Results of the cosmic tests. The filled and open circles
  represent lightyield from the near end for short ($<2000$~mm) and
  long ($>2000$~mm) strips, respectivly. The solid curve shows the
  results for all strips. The arrow indicates acceptance
  threshold value.}
\label {lightyield}
\end{figure}

To control strip quality during the production process each strip is
subjected to cosmic ray test, while the light yield caused by cosmic muons
passing trough the strip is measured. The triggers are located at
the middle of the strip and at both its ends. The results of the tests
are shown in Fig.~\ref{lightyield}. The quality of the strips is found
to be very good. The number of photo electrons from the end, nearest
to the SiPM, is shown by solid curve. Only strips with light
yield more than $33$~p.e. are accepted for the further
production. This number is almost $1.5$ times larger than original
goal~\cite{belle2tdr}.

The mass production of the detector components have been started in
ITEP in September 2012. Till the end of the September 2013 more than
$75\%$ is complete: 14075 strips are produced and tested, 13808
successfully passed the cosmic test; 875 segments are assembled. Three
(out of four) batches of the detector components are sent to KEK, the
last one will be sent in November.

\section{Acknowledgments}
The work is supported by the Russian Ministry of Education and Science
contracts 1366.2012.2, and 14.A12.31.0006; and President
grant MK-2877.2012.2.

\end{document}